\documentclass[lettersize,journal]{IEEEtran}
\usepackage{amsmath,amsfonts}
\usepackage{algorithmic}
\usepackage{algorithm}
\usepackage{array}
\usepackage[caption=false,font=normalsize,labelfont=sf,textfont=sf]{subfig}
\usepackage{textcomp}
\usepackage{stfloats}
\usepackage{url}
\usepackage{verbatim}
\usepackage{graphicx}
\usepackage{cite}
\begin{document}

\newcommand{\urss}[1]{\ensuremath{_{\mathrm{#1}}}}

\title{Direct Imaging of Transition-Edge Sensors with  Scanning SQUID Microscopy}

\author{Samantha Walker*, Austin Kaczmarek*, Jason Austermann, Douglas Bennett, Shannon M. Duff, Johannes Hubmayr, Ben Keller, Kelsey Morgan, Colin C. Murphy, Daniel Swetz, Joel Ullom, Michael D. Niemack, Katja C. Nowack
\thanks{Manuscript received September 25, 2024.

*S. W. and A. K. contributed equally to this work. \textit{(Corresponding author: Samantha Walker.)}

S. Walker is with the Department of Physics, Cornell University, Ithaca, NY, 14853 USA and the Cornell Center for Materials Research, Ithaca, NY, 14853 USA (e-mail: swalker@cornell.edu).

A. Kaczmarek, B. Keller, C. Murphy, and K. C. Nowack are with the Department of Physics, Cornell University, Ithaca, NY, 14853 USA.

M. D. Niemack is with the Department of Physics and the Department of Astronomy, Cornell University, Ithaca, NY, 14853 USA.

J. Austermann, D. Bennett, S. M. Duff, J. Hubmayr, and D. Swetz are with NIST, Boulder, CO, 80305 USA.

K. Morgan and J. Ullom are with NIST, Boulder, CO, 80305 USA and the Department of Physics, University of Colorado Boulder, Boulder, CO, 80302 USA.}}

\markboth{IEEE Transactions on Applied Superconductivity, ~Vol.~X, No.~X, September 2024}
{Shell \MakeLowercase{\textit{et al.}}: A Sample Article Using IEEEtran.cls for IEEE Journals}


\maketitle

\begin{abstract}
Significant advancements have been made in understanding the physics of transition-edge sensors (TESs) over the past decade. However, key questions remain, particularly a detailed understanding of the current-dependent resistance of these detectors when biased within their superconducting transition. We use scanning superconducting quantum interference device (SQUID) microscopy (SSM) to image the local diamagnetic response of aluminum-manganese alloy (Al-Mn) transition-edge sensors (TESs) near their critical temperature of approximately 175\,mK. By doing so, we gain insights into how the device dimensions influence TES transition width, which in turn affects device operation and informs optimal device design. Our images reveal that the Al-Mn thin film near the niobium (Nb) leads exhibits an excess diamagnetic response at temperatures several milli-Kelvin (mK) higher than the bulk of the film farther from the contacts. A possible origin of this behavior is a longitudinal proximity effect between the Nb and Al-Mn where the TES acts as a weak link between superconducting leads. We discuss how this effect shapes the temperature dependence of the resistance as the spacing between the leads decreases. This work demonstrates that magnetic imaging with SSM is a powerful tool for local characterization of superconducting detectors. 
\end{abstract}

\begin{IEEEkeywords}
superconducting, detector, transition-edge sensor, TES, bolometer, imaging, SQUID, scanning SQUID microscopy, diamagnetism, proximity effect.
\end{IEEEkeywords}

\section{Introduction}
\IEEEPARstart{S}{uperconducting} detectors have been of immense interest to the astronomy community over the past twenty years. These detectors utilize phenomena in superconductivity to enable photon-noise-, or background-noise-, limited performance and good optical efficiency. These types of detectors have demonstrated powerful capabilities for observing electromagnetic radiation across wavelengths spanning many orders of magnitude, from millimeter (mm) to gamma-ray wavelengths. The transition-edge sensor (TES) bolometer is a highly sensitive and stable superconducting detector used to measure incident radiation through heating of a strongly temperature-dependent resistor \cite{Irwin2005}, in this case a superconducting film biased in its superconducting transition. Through voltage-biasing the TES for negative electrothermal feedback, device operation is well-stabilized in the superconducting transition, even after power absorption. Large arrays of these detectors have significantly advanced our understanding of the Universe’s history and its contents through measurements of the cosmic microwave background (CMB) at mm wavelengths (e.g. \cite{Niemack2010,Filippini2010,Arnold2012,Austermann2012,BICEP22015,Henderson2016,anderson2018,Hui2018,SO2019, Dahal2020,Choi2020}).

Significant advancements have been made in our understanding of the physics of TESs over the past decade (e.g. \cite{Sadleir2010,Kozorezov2011,Bennett2012,Swetz2012,Bennett2013,Morgan2017}). However, key questions remain, particularly a detailed understanding of the current-dependent resistance of these detectors when voltage-biased in their superconducting transition (e.g. \cite{Sadleir2010,Kozorezov2011,Irwin1998,Bennett2012}). To date, TES devices have typically been characterized through electrical measurements \cite{Ullom2015}, including resistance ($R\urss{TES}$) vs bath temperature ($T\urss{b}$) curves and current-voltage ($I$-$V$) curves.  

In this work, we gain insights into how the spacing between the niobium (Nb) leads in aluminum-manganese alloy (Al-Mn) TES bolometers \cite{Deiker2004,Li2016,Duff2016} influences the width and critical temperature $T\urss{c}$ of the TES resistive transition. Transition width has been found to affect optimal device design, operation, and noise \cite{Ullom2015}. 
Using scanning superconducting quantum interference device (SQUID) microscopy (SSM), we directly image the local diamagnetic response of TESs with micrometer scale resolution \cite{Kirtley2010} and correlate these images with resistance measurements. This approach offers a new way to look at these devices which we show can provide valuable insight into device operation. 


\section{Experimental Setup}
We operate our SSM in a cryogen-free dilution refrigerator with a base temperature of approximately 10\,mK\cite{Low2021}. In our SSM, the SQUID features a 0.75\,$\mu$m radius pickup loop and a concentric field coil with radius 3\,$\mu$m which are scanned close to the surface of a sample \cite{Huber2008}. To detect the local diamagnetic response of a superconducting sample, we apply an AC current at a low frequency to the field coil, and use lock-in detection to measure the induced magnetic flux in the pickup loop. Additionally, we use the DC flux through the pickup loop to image superconducting vortices and as a readout by detecting the magnetic field that the TES bias current produces.

The sample is an early prototype Al-Mn TES bolometer chip, which contains 12 TES devices originally designed for the Advanced Atacama Cosmology Telescope \cite{Henderson2016,Choi2020} and fabricated in the NIST Boulder Microfabrication Facility. We measure three TES bolometers on this chip with different spacing between Nb leads ($L$) and varying device widths ($W$). Device a is the TES with the largest $L$, with dimensions $L=25$\,$\mu$m, $W=250$\,$\mu$m, a 1/10 aspect ratio, and consequently a resistance given by 1/10\,$R\urss{s}$ where $R\urss{s}$ is the square resistance of the Al-Mn film. Device b has a 1/5 aspect ratio with dimensions $L=20$\,$\mu$m, $W=100$\,$\mu$m and a resistance of 1/5\,$R\urss{s}$. Device c has a 1/10 aspect ratio with dimensions $L=10$\,$\mu$m, $W=100$\,$\mu$m and a resistance of 1/10\,$R\urss{s}$. Devices b and c have a smaller $W$ compared to Device a, although Devices a and c are designed to have an identical normal resistance $R\urss{n}$ and aspect ratio. In addition, Device a has an aspect ratio and Nb lead spacing similar to TES bolometers used in CMB experiments, specifically a 1/8 aspect ratio with dimensions $L=25$\,$\mu$m, $W=200$\,$\mu$m and 1/8\,$R\urss{s}$ \cite{Henderson2016}.

\begin{figure}[t!] 
\centering
\includegraphics[width=3.4in]{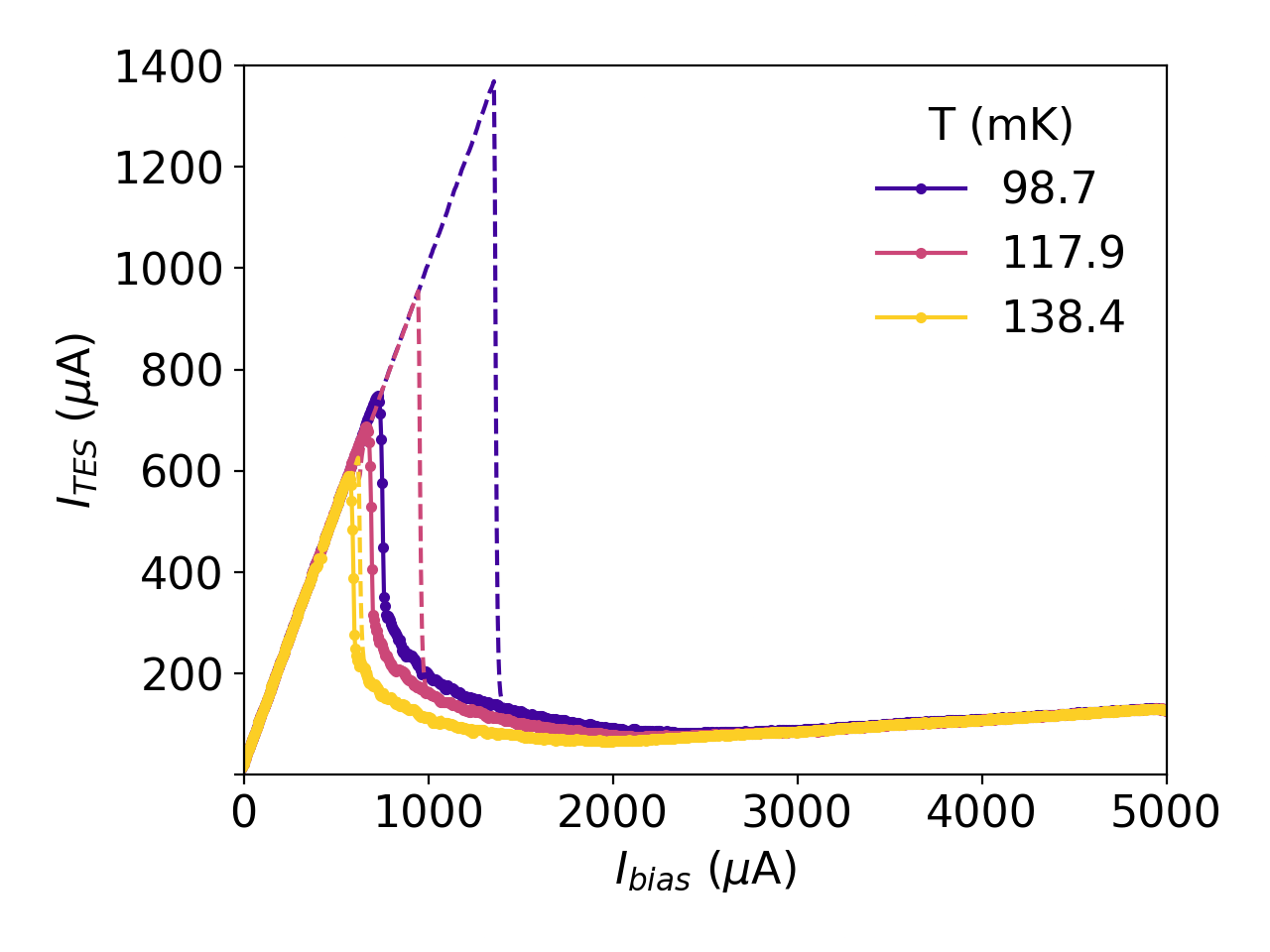} 
\caption{$I\urss{TES}$ vs. $I\urss{bias}$ curves for Device a with 1/10\,$R\urss{s}$ geometry measured at three bath temperatures, $98.7$\,mK, $117.9$\,mK, and $138.4$\,mK. In the dashed (solid) curves $I\urss{bias}$ is swept upwards (downwards). In this measurement, a scanning SQUID is used to inductively read-out the current flowing across the TES by positioning the SQUID within micrometers of the TES lead. 
\label{fig_1}}
\end{figure}

Each TES is voltage-biased through a parallel shunt resistor $R\urss{sh} \approx 180$\,$\mu\Omega$. However, unlike in a typical TES circuit \cite{Irwin2005}, an inductively coupled SQUID and SQUID array amplifier are not directly embedded in the circuit for readout. Instead, we measure the current through the TES $I\urss{TES}$ by positioning the scanning SQUID near the Nb lead connecting each TES bolometer and inductively couple the magnetic field produced by the TES bias current into the scanning SQUID. At a fixed position, the SQUID voltage signal $V\urss{SQ}$ and the TES current are linearly related. As we change the total bias current $I\urss{bias}$ applied to the circuit, we read-out the current in the TES similar to using a SQUID array amplifier. Figure\,\ref{fig_1} shows $I\urss{TES}$ vs. $I\urss{bias}$ curves measured using the scanning SQUID for Device a measured at three bath temperatures, $98.7$\,mK, $117.9$\,mK, and $138.4$\,mK, well below the critical temperature $T\urss{c}$ of this device. In the dashed curves, $I\urss{bias}$ is swept upwards, showing the TES transition to the normal branch after reaching the temperature-dependent critical current $I\urss{c}$. For the solid curves, $I\urss{bias}$ is swept from high current downward similar to typical TES voltage-biased operation. We assume that at low current $I\urss{TES} = I\urss{bias}$ and use this to calibrate the proportionality factor between the SQUID signal and $I\urss{TES}$. This is a reasonable assumption given that all connections within the bias circuit are superconducting including aluminum wirebonds. The shape of $I\urss{TES}$ vs. $I\urss{bias}$ curves we find is similar to TESs measured using a typical TES read-out circuit, validating our approach.

\section{Measurements and Results}
\subsection{TES resistance measurements}

\begin{figure}[t!] 
\centering
\includegraphics[width=3.1in]{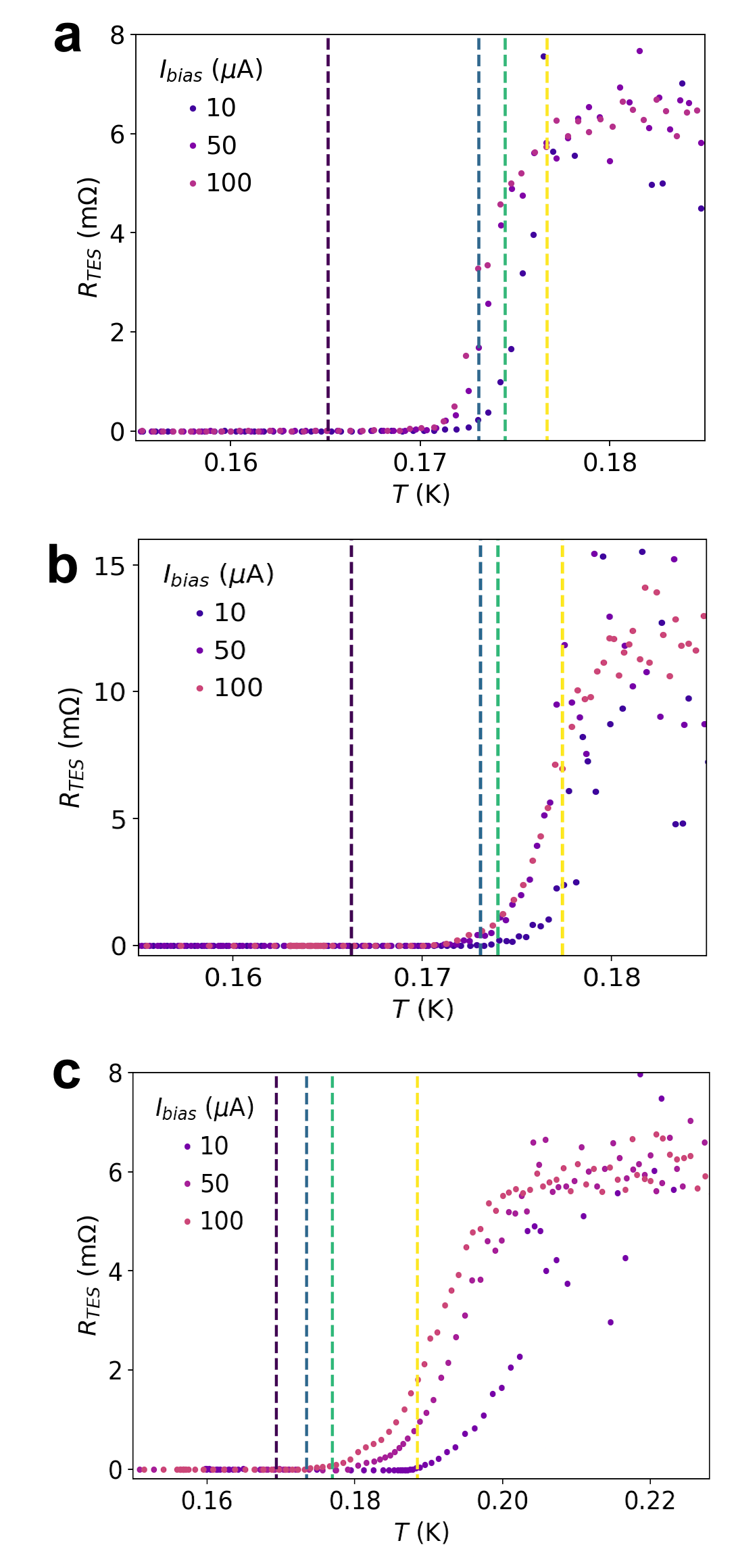} 
\caption{TES resistance vs bath temperature measurements for Devices a--c (top to bottom) arranged from largest to smallest Nb lead spacing, 25\,$\mu$m, 20\,$\mu$m, and 10\,$\mu$m, respectively. Different colored curves correspond to different values of bias current $I\urss{bias}$, where $R\urss{TES} = V\urss{TES}/I\urss{TES}$ is measured using the SSM at a fixed position near the Nb lead for each device. Vertical dashed lines correspond to the temperatures of the magnetic susceptibility $\delta M$ images shown in Fig.\,\ref{fig_3}. Compared to the other TES bolometer geometries, Device c shows an extended superconducting transition, higher $T\urss{c}$, and $T\urss{c}$ dependence with $I\urss{bias}$, consistent with the proximity effect between the Nb leads and Al-Mn TES.}
\label{fig_2}
\end{figure}

We characterize the temperature dependence of the TES resistance $R\urss{TES}$ at different bias currents $I\urss{bias}$ using the same method as in Fig.\,\ref{fig_1}. We refer to the $T\urss{c}$ for each TES in this work as the bath temperature where the TES first starts to become resistive at very low bias current $I\urss{bias}=10$\,$\mu$A. 
As expected, we observe that $R\urss{TES}$ changes with applied $I\urss{bias}$. 
To measure $R\urss{TES}$-$T$ curves, we sweep the bath temperature continuously and slowly. At each temperature, we successively measure the SQUID signal of the voltage-biased TES with the bias current on, $V\urss{SQ}(I\urss{bias})$, and with the bias current off, $V\urss{SQ}(I\urss{bias}=0)$. 
The difference in the SQUID signal $\Delta V\urss{SQ} = V\urss{SQ}(I\urss{bias}) - V\urss{SQ}(I\urss{bias}=0)$ is insensitive to arbitrary offset voltages in the SQUID readout, which are always present. The TES current is then related linearly to the difference in the SQUID signal, $I\urss{TES}=\gamma\Delta V\urss{SQ}$, where $\gamma$ is determined by assuming $I\urss{TES}=I\urss{bias}$ well below $T\urss{c}$ as above. The TES voltage $V\urss{TES}$ is equal to the voltage across the shunt resistor $V\urss{TES} = R\urss{sh}I\urss{sh} = R\urss{sh}(I\urss{bias}-I\urss{TES})$. This allows us to calculate the TES resistance by $R\urss{TES} = V\urss{TES}/I\urss{TES}$. For our measurement, the extracted $R\urss{TES}$ is noisy in the normal state. This is a result of the conversion process from $\Delta V\urss{SQ}$ to $R\urss{TES}$. Although $\Delta V\urss{SQ}$ is not nearly this noisy and is well above the noise floor, the conversion shifts the measured $\Delta V\urss{SQ}$ to a value of $I\urss{TES}$ that is very small in the normal state. Dividing by this small value of $I\urss{TES}$ to calculate $R\urss{TES}$ amplifies any fluctuations.


Fig.\,\ref{fig_2} shows $R\urss{TES}$-$T$ curves for Devices a--c. Curves correspond to different values of bias current $I\urss{bias}$ as indicated. We find that Device c, with the closest spacing between Nb leads, shows a broader transition, spanning about $25$\,mK compared to about $5$\,mK to $8$\,mK in Devices a and b. In addition, Device c shows the strongest dependence on $I\urss{bias}$ where the TES first becomes resistive. These effects are consistent with the proximity effect between the Nb leads and Al-Mn film affecting the superconducting transition of Device c.

The $R\urss{TES}$-$T$ curves presented here are consistent with previous measurements of devices with the same dimensions and from the same fabrication run. In those measurements, the TES bolometers were voltage-biased and inductively coupled to a SQUID and SQUID array amplifier. We find comparable normal resistance values $R\urss{n}\sim6.5$\,m$\Omega$ for Device\,a, and $R\urss{n}\sim13$\,m$\Omega$ for Device b and Device c. 
We also find consistent logarithmic sensitivity of the superconducting transition $\alpha = d\log R/d\log T$ values at small bias currents. 
For $I\urss{bias}=100$\,$\mu$A, we calculate the mean alpha from the transition width for each TES to be $\alpha \sim 148$ for Device a, $\alpha \sim 104$ for Device\,b, and $\alpha \sim 52$ for Device\,c.



\subsection{Imaging the local diamagnetic susceptibility}

\begin{figure*}[!t]
\centering
\captionsetup[subfigure]{labelformat=empty}
\subfloat[]{\includegraphics[width=1.67in]{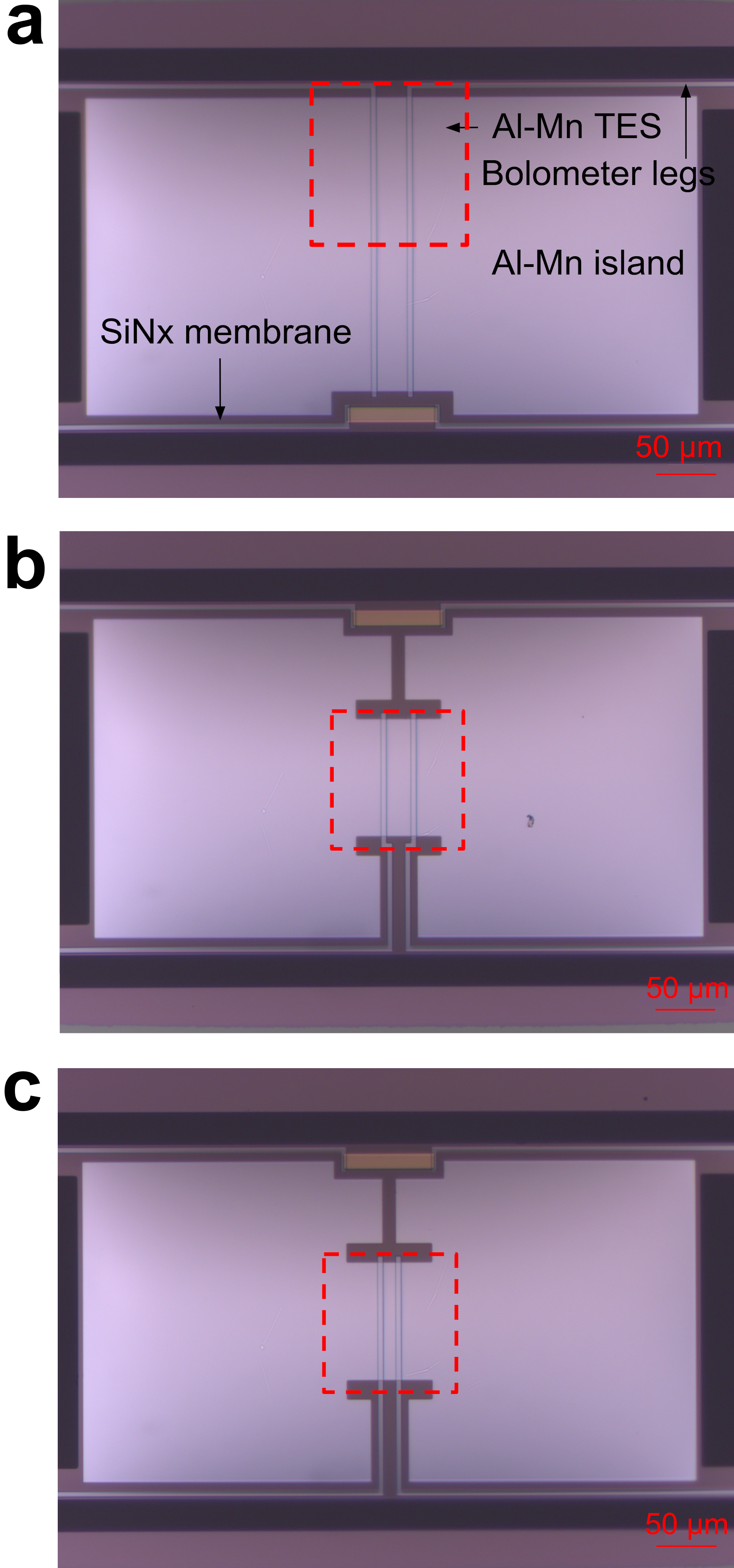}%
\label{fig_3left}}
\hfil
\subfloat[]{\includegraphics[width=5.1in]{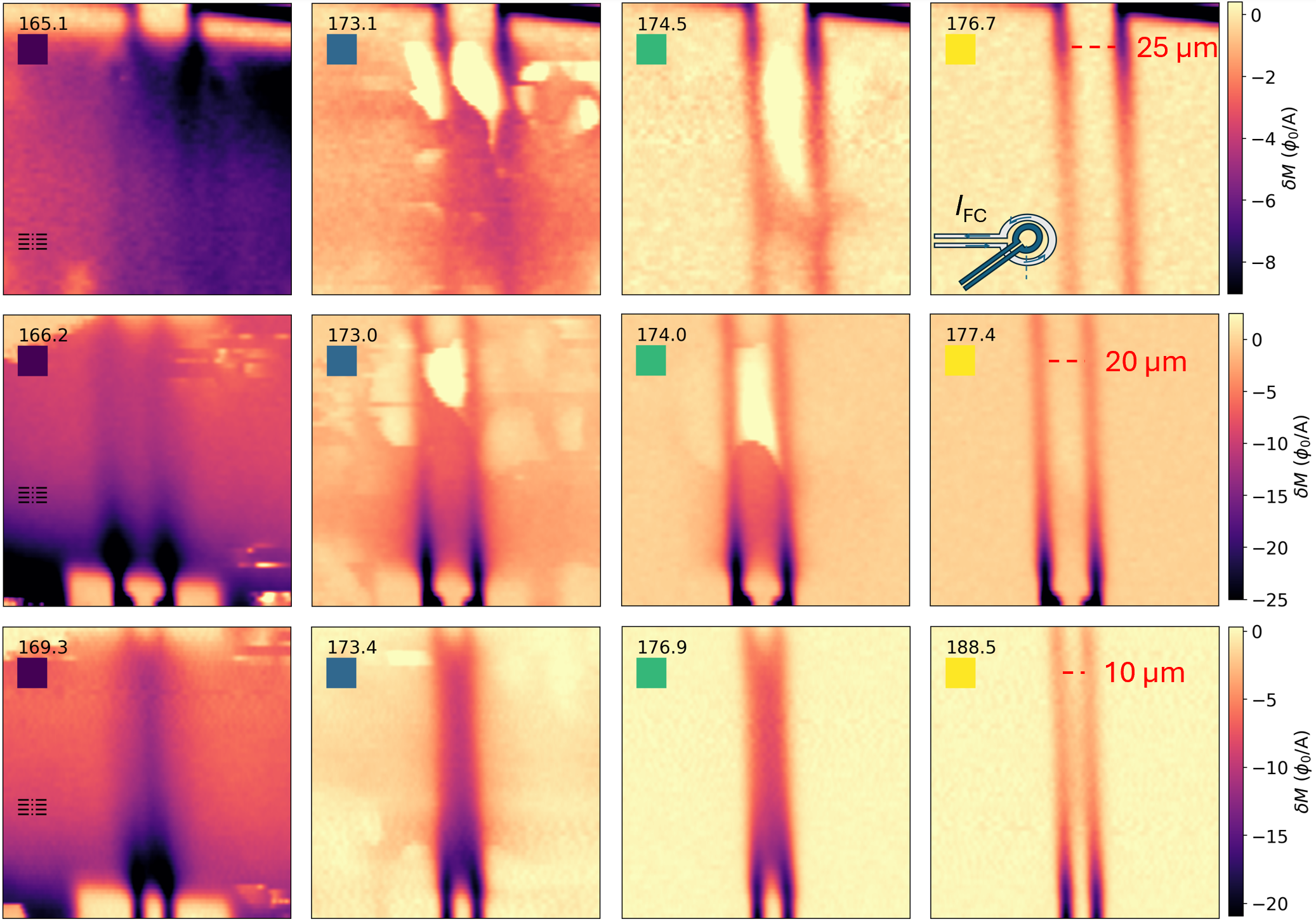}%
\label{fig_3right}}
\caption{Device a, b, and c (top to bottom) arranged from largest to smallest Nb lead spacing. \textit{Left column:} Optical images of each Al-Mn TES bolometer with Nb leads (5\,$\mu$m trace width) suspended on a silicon nitride (SiNx) membrane. Important parts are labeled. The red dashed box denotes the region imaged with SSM. \textit{Right four columns:} From left to right, SSM images of the local diamagnetic response of each Al-Mn TES well below $T\urss{c}$ (left), below and near $T\urss{c}$ (middle two), and above $T\urss{c}$ (right). The Nb leads are superconducting in all images. The lead spacing is annotated with a red dashed line and text in the rightmost image. The temperature each image was measured at is labeled to the upper left and the small colored squares correspond to the vertical dashed lines in Fig.\,\ref{fig_2}. The small black dot-dashed lines denote positions used for linecuts in Fig.\,\ref{fig_4}. Each row of SSM images are displayed on the same color scale, in which darker colors correspond to larger local diamagnetic response $|\delta M|$. We include a cartoon schematic of the scanning SQUID, including the field coil current $I\urss{FC}$, in the upper rightmost image.}
\label{fig_3}
\end{figure*}

In Figure\,\ref{fig_3} left column, we show optical micrographs of Al-Mn TES bolometers with Nb leads suspended on a silicon nitride (SiNx) membrane. Device a, b, and c are arranged largest to smallest Nb lead spacing from top to bottom, 25\,$\mu$m, 20\,$\mu$m, and 10\,$\mu$m, respectively. We label some important parts of each TES bolometer and include a red dashed box to denote the approximate region that was imaged with the SSM. 

We probe the local magnetic response by running an AC current through the field coil of about 250\,$\mu$A at a frequency of 270\,Hz. Far from the sample, this measurement yields the mutual inductance, $M=M_0$, between the pickup loop and field coil. Close to a superconductor, the sample generates currents to screen the local magnetic field produced by the current in the field coil. This effectively reduces the mutual inductance $M$. The change, $\delta M = M-M_0$, therefore allows us to detect superconducting regions in the device where $\delta M<0$. The measured response is directly related to the London penetration depth and the superfluid density, although we do not quantitatively evaluate this in this work.

We do not use any magnetic shielding, but instead use a Helmholtz coil to cancel the z-component of any background magnetic field. Prior to imaging the diamagnetic susceptibility, we adjust the compensation field and cool the TES through $T\urss{c}$, until we find the compensation field that minimizes the number of superconducting vortices in the Al-Mn film by directly imaging them with the SSM. This scheme is not perfect, and we observe the influence of vortices in some of our susceptibility images. We believe the residual vortices are due to some warping of the SiNx membrane that allows in-plane fields to couple out-of-plane flux through the Al-Mn film which induces vortices.




\begin{figure}[t!]
\centering
\includegraphics[width=3.08in]{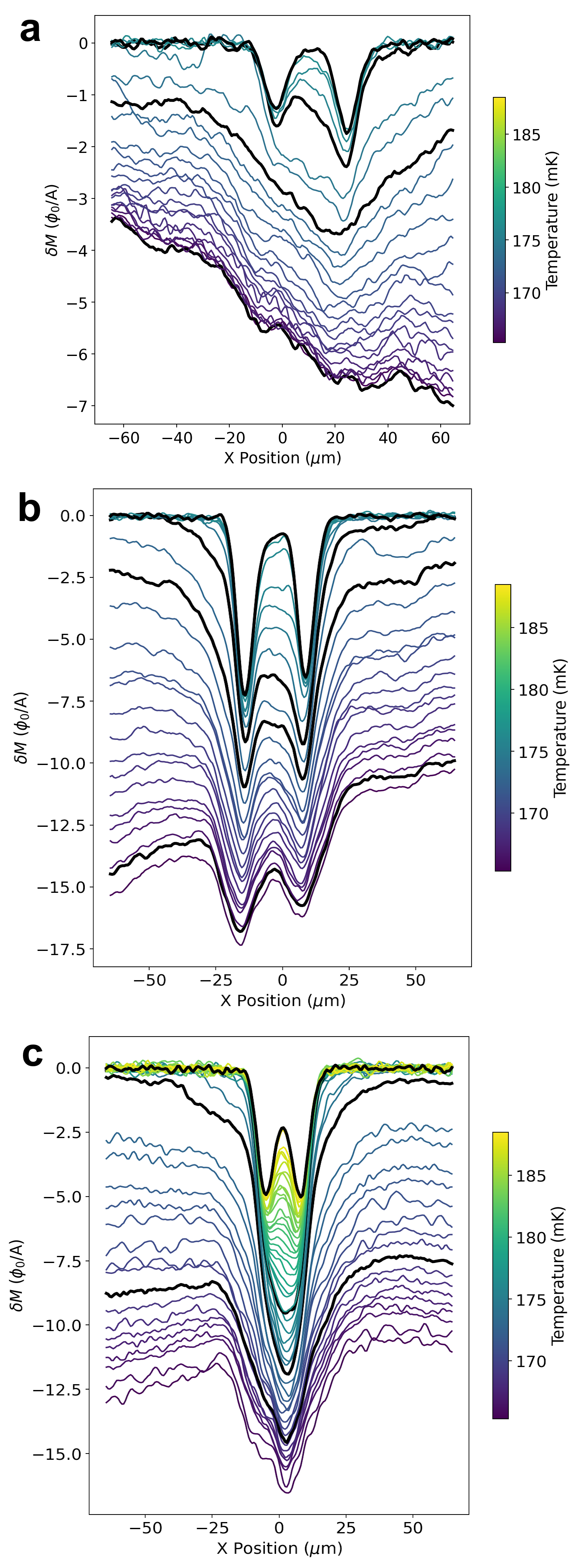} 
\caption{Horizontal linecuts in susceptibility images as a function of temperature as in Fig.\,\ref{fig_3} for Devices a--c (top to bottom). Linecuts are the average of four lines denoted by the dot-dashed lines in Fig.\,\ref{fig_3}. Linecuts corresponding to images in Fig.\,\ref{fig_3} are highlighted by bold black curves. In each device, the dips at higher $T\urss{b}$ correspond to the position and diamagnetic response from only the Nb.}
\label{fig_4}
\end{figure}

Figure\,\ref{fig_4} shows images of the local diamagnetic response for Devices a--c acquired using SSM. We change the bath temperature near the $T\urss{c}$ of each device. The images are acquired with no bias current applied to the TES circuit. The different columns show SSM images of the local diamagnetic response measured at bath temperatures well below $T\urss{c}$ (left), below and near $T\urss{c}$ (middle two), and above $T\urss{c}$ (right). The temperatures are labeled in each image and the colored square corresponds to the temperature of the same color dashed line in Fig.\,\ref{fig_2}. 
Each row of SSM images is displayed on the same color scale, in which darker colors correspond to a larger local diamagnetic response $|\delta M|$ in units of $\phi_0$/A. We see some imaging artifacts, for example, variations in the signal strength from the Nb leads at the highest temperatures. We suspect these are caused by variations of the scan height throughout the image likely due to the warping of the SiNx membrane mentioned above.

At bath temperatures below and near the superconducting transition, we observe an enhanced local diamagnetic response between the Nb leads and Al-Mn TES in each device geometry measured. In addition, we observe in some images irregular and sharp signals which we believe originate from the dynamics of residual vortices and flux trapping in the devices (e.g. the bright lobe feature in Device a at 173.1\,mK). In the following discussion, we focus on the smoother, qualitative features in the images. Specifically, we observe a region of higher magnetic susceptibility emanating from the Nb leads out into the Al-Mn film. The strength and length scale of this spreading diamagnetism increases as temperature is lowered and eventually results in some amount of overlap between the leads. In Device c, with the smallest spacing between the Nb leads, we observe an enhanced local diamagnetic response that appears to spread uniformly between the Nb leads and Al-Mn bolometer at higher bath temperatures than in other device geometries and persists at temperatures well above the $T\urss{c}$ of the bare Al-Mn film. This suggests that the Nb leads are proximitizing the Al-Mn film, raising the local $T\urss{c}$ and superfluid density of the Al-Mn.


  

To study this proximity effect in further detail, Fig.\,\ref{fig_4} shows linecuts along the horizontal direction in images at more closely spaced temperatures. The black curves correspond to linecuts of images in Fig.\,\ref{fig_3}. To obtain the linecuts, we average over four lines in the images whose vertical positions we mark in Fig.\,\ref{fig_3}. Each plot shows dips at the highest temperatures that correspond to the local diamagnetic response from only the Nb leads. The Nb leads are 5\,$\mu$m wide, and the diamagnetic response appears as a peak with a full width at half maximum of approximately 7\,$\mu$m to 8\,$\mu$m, providing a measure of the spatial resolution of our imaging. For Device c, we observe a slight enhancement of signal even at higher temperatures due to the overlap of the two peaks corresponding to the Nb leads.
As the bath temperature decreases, a diamagnetic response gradually emerges, initially appearing as a tail extending outward and growing from the Nb leads. Simultaneously, the response grows in the region between the Nb leads. Away from the leads, the response initially remains zero as long as the temperature is above the critical temperature $T\urss{c}$ of the bare Al-Mn film, where here $T\urss{c}$ refers to the temperature at which the bare Al-Mn film first shows a diamagnetic response. Below $T\urss{c}$, a signal that increases with decreasing bath temperature also appears away from the Nb leads. Although a quantitative analysis is beyond the scope of this work, we note that the signal strength can be related to the London penetration depth \cite{Kirtley2010}. The rise in signal strength away from the Nb leads reflects that the London penetration of the Al-Mn film is decreasing over the temperature range that we explore. 

The data implies that there is a temperature range where the region between the Nb leads remains superconducting, even when the temperature is above the $T\urss{c}$ of the bare Al-Mn film. This effect becomes increasingly pronounced as the lead spacing decreases from Devices a to c. Device c with the smallest lead spacing exhibits a significantly stronger proximity effect than the other geometries, showing a local diamagnetic response in between the leads at temperatures more than $10$\,mK higher than the $T\urss{c}$ of the bare Al-Mn. This observation aligns with our $R\urss{TES}$-$T$ measurements (Section\,IIIB) where Device c has a much higher $T\urss{c}\sim190$\,mK compared to the other TES devices ($T\urss{c}\sim175$\,mK) and a broadened transition, from approximately $5$\,mK to $25$\,mK.
This broadening of the TES transition correlates with a significantly lower mean $\alpha$ value for Device c, $\alpha\sim 52$, compared to $\alpha\sim 104$ for Device b, and $\alpha\sim 148$ for Device a. Lower $\alpha$ values are directly linked to lower TES loop gain values and longer time constants \cite{Irwin2005}. 

\section{Conclusions}
In this work, we use SSM to directly image the superconducting transition in Al-Mn TESs to gain insights into the behavior of this resistive transition. 
We study the enhanced local diamagnetism between the Nb leads and Al-Mn TES bolometers of different geometries, and correlate these results with measurements of the width and $T\urss{c}$ of the TES resistive transition. Our findings reveal that as device lead spacing decreases, both the transition width and $T\urss{c}$ increase, as observed in measurements of the resistance and the local diamagnetism. This leads to significantly lower mean $\alpha$ values in the transition, where alpha is linearly related to TES loop gain values and results in longer device time constants \cite{Irwin2005}. These findings provide evidence of the proximity effect between the Nb leads and the Al-Mn film, which can significantly impact device performance and should be considered in device design. 
The role of the proximity effect has been inferred from electrical characterization of TES devices \cite{Sadleir2010}, but this work provides the first direct observation of this effect through imaging. 

In future work, we plan to use SSM to correlate device behavior at lower bath temperatures when biasing the TESs at various points in their transitions. In addition to imaging the local diamagnetic susceptibility, we plan to also image the spatial distribution of current flow inside TESs when voltage-biased at different percentages of normal resistance. This study demonstates that magnetic imaging with SSM is a powerful tool for locally characterizing the diamagnetism in TESs, offering a more detailed understanding of these detectors and providing guidance for optimizing their performance. 


\section*{Acknowledgments}
S. Walker acknowledges support from the Cornell University Research Excellence Scholars (CURES) Fellowship and the Cornell Center for Materials Research. This work was partially funded by the Cornell Center for Materials Research through NSF MRSEC program grant DMR-1719875. A. Kaczmarek acknowledges support from the Air Force Research Laboratory, Project Grant AFOSR grant FA9550-21-1-0429.





\clearpage

\bibliography{main.bbl} 
\bibliographystyle{IEEEtran}
 
\end{document}